\documentstyle[12pt]{article}
\newcommand{\be}{\begin{equation}}
\newcommand{\ee}{\end{equation}}
\newcommand{\ba}{\begin{eqnarray}}
\newcommand{\ea}{\end{eqnarray}}
\newcommand{\bb}{\begin{thebibliography}}
\newcommand{\eb}{\end{thebibliography}}
\newcommand{\ci}[1]{\cite{#1}}

\begin{document}
\begin{center}
{\Large\bf { THE $\tau^- \to (\omega,\phi) \pi^- \nu_{\tau}$ DECAYS IN
                     THE CHIRAL MODEL }}\\
\vspace*{0.5cm}
{\large  K. R. Nasriddinov, B. N. Kuranov, T. A. Merkulova  } \\
\end{center}
{\large
\begin{center}
{\ Institute of Nuclear Physics, Academy of Sciences of Uzbekistan,\\
    Ulugbek, Tashkent, 702132 Uzbekistan }\\
\end{center}
}
\begin{quote}
   The $ \tau^- \to (\omega,\phi) \pi^- \nu_{\tau}$ decays of the
$\tau$ lepton are studied using the method of phenomenological
$SU(3) \times SU(3)$ chiral Lagrangians. It is shown that
the obtained value for the $\tau^{-} \to \phi \pi^- \nu_{\tau}$ decay
probability is very sensitive to deviations from the mixing angle.
Calculated partial widths for these decays are compared with the
available experimental and theoretical data.\\
\end{quote}

{\large
At present the studies of the $ \tau^- \to (\omega,\phi) \pi^- \nu_
{\tau}$ decays are of great interest. These investigations could be
provide evidence axial vector second class currents \ci{pr1} and
useful information on the $\omega-\phi$ mixing angle.

The $ \tau^{-} \to \omega \pi^- \nu_{\tau}$ decay mode has been
studied in the framework of the vector-meson dominance model \ci{zp2},
a low energy $U(3) \times U(3)$ chiral Lagrangian model \ci{prd3} and,
using the heavy vector-meson chiral perturbation formalism \ci{prd4}.
The $ \tau^{-} \to (\omega,\phi) \pi^- \nu_{\tau}$ decay channels have
been studied using the CVC hypothesis \ci{plb5} and the vector meson
dominance model \ci{prd6} as well.

In this paper we study the $ \tau^{-} \to (\omega,\phi) \pi^- \nu_
{\tau}$ decay channels using the method of phenomenological
$SU(3) \times SU(3)$ chiral Lagrangians (PCL's) \ci{prl7}.
Recently \ci{prd8} the $ \tau \to VP \nu_{\tau}$ decay channels of
the $\tau$ lepton have been studied also using this method.
In order to investigate these decay
modes we have obtained the expression of weak hadron currents between
pseudoscalar and vector meson states by including the gauge fields of
these mesons in covariant derivatives \ci{pn9}. The weak hadron
currents in this way have the form
\vspace*{0.3cm}
\be
J^{i}_{\mu} = F_{\pi}gv^a_{\mu} \varphi^bf_{abi},
\ee
\vspace*{0.3cm}
where $F_{\pi}$ = 93 MeV, $g$ is the "universal" coupling constant,
which is fixed from the experimental $\rho \to \pi \pi$ decay width:
\vspace*{0.3cm}
$$ \frac{g^2}{4 \pi} \simeq3.2, $$
$v^a_{\mu}$ and $\varphi^b$ represent the fields of the $1^-$ and
$0^-$ mesons, respectively.

The partial widths of
the $ \tau^{-} \to (\omega,\phi) \pi^- \nu_{\tau}$ decay channels are
equal to zero according to the obtained expression of weak hadron
currents Eq.(1). Therefore these channels can be realized via effects
of secondary importance. Here, we calculate the partial widths of
these decay modes in the framework of the abovementioned method. Note
that the hadron decays of the $\tau$ lepton up to three pseudoscalar
mesons in the final state have been studied also in the framework of
this method \ci{ydf10,upj11}.

In the PCL's, the weak interaction Lagrangian has the form
\vspace*{0.3cm}
\be
L_W =\frac{G_F}{\sqrt{2}} {J^h_{\mu} l^+_{\mu} + h.c.},
\ee
\vspace*{0.3cm}
where  $G_{F}\simeq10^{-5}/m^{2}_{P}$ is the Fermi constant,\\
$ l_{\mu} = \bar {u_{l}}\gamma_{\mu}(1+\gamma_{5})u_{\nu_{l}}$ is the
lepton current, and hadron currents have the form \ci{prl7}
\vspace*{0.5cm}
$$ J^h_{\mu} = J^{1+i2}_{\mu} \cos{\Theta_c} + J^{4-i5}_{\mu} \sin
 {\Theta_c},$$
where $\Theta_c$ is the Cabibbo angle, and $\rho,\rho^{\prime}
...$ meson currents are defined as
$$ J^{1+i2}_{\mu}=\frac{m^2_{\rho}}{g} \rho^-_{\mu}. $$

The strong interaction Lagrangian of vector mesons with
vector and pseudoscalar mesons has the form
\be
L_s(vv \varphi) = -g_{vv \varphi} \varepsilon_{\mu \nu \alpha \beta}
Sp(\partial_{\mu} \hat V_{\nu} \partial_{\alpha} \hat V_{\beta}
\hat \varphi),
\ee
where $g_{vv \varphi}=3g^2/16 \pi^2 F_{\pi}$ is the coupling constant,
$ \hat V_{\mu}=\frac{1}{2i} \lambda_i v^i_{\mu}$, and \\
$ \hat \varphi=\frac{1}{2} \lambda_i \varphi^i $.

According to this equation the strong interaction
Lagrangians of the $\rho^-,\rho^{\prime-}...$ mesons with
($\omega,\phi$) and $\pi^-$ mesons, at $39^o$ of the $\omega-\phi$
mixing angle, have the forms
\be
L_S(\rho^- \to \omega \pi^-) = -\frac{i}{2}g_{vv \varphi}
 \varepsilon_{\mu \nu \alpha \beta} \partial_{\mu} \omega_{\nu}
 \partial_{\alpha} \rho^+_{\beta} \pi^-,
\ee
\be
L_S(\rho^- \to \phi \pi^-) = -0.0016ig_{vv \varphi}
 \varepsilon_{\mu \nu \alpha \beta} \partial_{\mu} \phi_{\nu}
 \partial_{\alpha} \rho^+_{\beta} \pi^-.
\ee

According to the Lagrangians (2) and (4) the decay amplitude for the
$ \tau^{-} \to \omega \pi^- \nu_{\tau}$ decay channel has the form \\
\be
M =\frac{G_Fm^2_{\rho}g_{vv \varphi}cos{\Theta_c}}{\sqrt{8}g[S_1-m^2
 _{\rho}-i(m_{\rho} \Gamma_{\rho})]}[\varepsilon_{\mu \nu \alpha
 \beta}(P^{\omega}_{\alpha} K^{\rho}_{\mu} \epsilon^{\lambda}_{\beta}
 (\omega)) \bar {u_{\nu}}\gamma_{\nu}(1+\gamma_{5})u_{\tau}],
\ee
where $\epsilon^{\lambda}_{\beta}(\omega)$ is the polarization vector
of $\omega$-meson, $P^{\omega}_{\alpha}$, $K^{\rho}_{\mu}$ are the
4-momenta of the $\omega$- and $\rho$-mesons,respectively.
Using Eq.(6) we have defined the squared matrix element of this
process\\
$ \mid M \mid^2=-\\
K[m^2_{\omega}m^2_{\pi}(0.5(S_2-m^2_{\pi})+0.5(S_3-m^2_{\omega}))-m^2
_{\omega}(0.5(S_2-m^2_{\pi}))^2- \\
  m^2_{\pi}(0.5(S_3-m^2_{\omega}))^2+0.25(S_2-m^2_{\pi})(S_3-m^2_
	{\omega})(S_1-m^2_{\pi}-m^2_{\omega})-  \\
  (0.5(S_1-m^2_{\pi}-m^2_{\omega}))^2(0.5(S_2-m^2_{\pi})+0.5(S_3-m^2_
	{\omega}))] $,  \\

\vspace*{0.3cm}
where
$$ K=\frac{2G^2_Fm^4_{\rho}(3gcos{\Theta_c})^2}
   {(16 \pi^2F_{\pi})^2[(S_1-m^2_{\rho})^2+(m_{\rho} \Gamma_
	 {\rho})^2]},$$
\vspace*{0.3cm}
$S_1$,$S_2$ and $S_3$ are the Mandelstam variables. Note that the
squared matrix element for the $ \tau^{-} \to \phi \pi^- \nu_{\tau}$
decay channel can be easily obtained by replacing
$m_{\omega} \to m_{\phi}$ and the corresponding $K$, which
is defined according to the Lagrangian (5).

Using Lagrangians (2), (4), and (5) we calculated the partial width of
the $ \tau^{-} \to (\omega,\phi) \pi^- \nu_{\tau}$ decay channels
by means of the TWIST code \ci{rep12} and obtained for the first decay
channel
$\Gamma(\tau^{-} \to \omega \pi^- \nu_{\tau})=0.44 \times 10^{11}
sec^{-1} $.\\
This obtained result for the partial width is consistent
(within errors) with the experimental value
$\Gamma(\tau^{-} \to \omega \pi^- \nu_{\tau})=(0.54 \pm 0.17) \times
10^{11} sec^{-1} $ \ci{prd13} \\
and with the VMD model prediction
$\Gamma(\tau^{-} \to \omega \pi^- \nu_{\tau})=(0.42 \pm 0.19) \times
10^{11} sec^{-1} $ \ci{prd6} than the prediction \ci{prd3}
$\Gamma(\tau^{-} \to \omega \pi^- \nu_{\tau})=0.31 \times 10^{11}
sec^{-1} $, but it lies below the prediction by the CVC hypothesis
$\Gamma(\tau^{-} \to \omega \pi^- \nu_{\tau})=(0.73 \pm 0.10) \times
10^{11} sec^{-1} $ \ci{plb5}.
In these calculations we use $\omega-\phi$ mixing as
\begin{center}
$\omega=V_8sin{\Theta_V}+V_0cos{\Theta_V},$ \\
$\phi=V_8cos{\Theta_V}-V_0sin{\Theta_V},$ \\
\end{center}
and at the $\Theta_V=39^o$ for the probability of the
$\tau^{-} \to \phi \pi^- \nu_{\tau}$ decay channel we obtain
\begin{center}
$\Gamma(\tau^- \to \phi \pi^- \nu_{\tau})=0.38 \times 10^{6} s^{-1}$.\\
\end{center}
This obtained result is consistent with the experimental data
\ci{prd14} $\Gamma(\tau^- \to \phi \pi^- \nu_{\tau})<(12.04 \pm 0.07)
\times 10^{8} s^{-1}$, but it lies below the VMD model \ci{prd6}
prediction $\Gamma(\tau^- \to \phi \pi^- \nu_{\tau})=(0.41 \pm 0.17)
\times 10^{8} s^{-1} $,
and almost four orders of magnitude below the upper limit obtained
using the CVC hypothesis \ci{plb5}
$\Gamma(\tau^- \to \phi \pi^- \nu_{\tau})<0.31 \times 10^{10} s^{-1}$.
And at the ideal mixing angle $\Theta_V=35,3^o$, for this decay
channel we obtain
$\Gamma(\tau^- \to \phi \pi^- \nu_{\tau})=1.27 \times 10^{8} s^{-1} $,
which is lies above the prediction \ci{prd6}. We can observe that the
obtained value for the $\tau^{-} \to \phi \pi^- \nu_{\tau}$ decay
probability is very sensitive to deviations from the mixing angle as
$$\frac{1}{2\sqrt{3}}cos{\Theta_V}-\frac{1}{2\sqrt{2}}sin{\Theta_V}.$$\\
Therefore this decay channel could be used as ideal
"source" for the $\omega-\phi$ mixing study and could be defined
from the measured partial width of the $\tau^- \to \phi \pi^- \nu_
{\tau}$ decay channel at a $\tau$-charm factory with high accuracy.
Note that the $\tau^{-} \to \omega \pi^- \nu_{\tau}$ decay
probability is almost independent from the $\omega-\phi$ mixing
angle (at the $\Theta_V=39^o$ and $\Theta_V=35,3^o$ is equals
approximately to 1/2) according to
$$\frac{1}{2\sqrt{3}}sin{\Theta_V}+\frac{1}{2\sqrt{2}}cos{\Theta_V}.$$\\

In these decays $\rho(1450)$- and $\rho(1700)$-vector intermediate
meson state contributions have been taken into account also
according
to Eqs. (2), (4), and (5), which have the same flavor quantum numbers
as $\rho(770)$ meson. Therefore
in our calculations we used also the same $g$-coupling constant,
as in Ref. \ci{prd8}, for
all the $\rho(770)$-, $\rho(1450)$-, and $\rho(1700)$-
vector intermediate meson states which have widths of 151, 310, and
235 MeV, respectively. And in these decays contributions of the
$\rho(1450)$- and $\rho(1700)$-vector intermediate meson states
dominate those of the $\rho(770)$ ones. Note
that in calculations \ci{prd3,prd4} contributions from these
higher radial excitations have not been taken into account. And in
Ref's \ci{zp2,prd6} these decay channels are calculated
assuming only one $\rho^{\prime}$ resonance in addition to the $\rho$
meson.

In summary, the Lagrangians (2), (4), and (5) allow us to describe of
$ \tau^{-} \to (\omega,\phi) \pi^- \nu_{\tau} $ decays in satisfactory
agreement with the available experimental data and theoretical
predictions. And, probably, taking into account corresponding coupling
constants, as in Ref. \ci{prd6}, would allow us to describe these
decays more correctly compared to these calculations.

We would like to express our deep gratitude to F.A.Gareev, F.Khanna,
Chueng-Ryong Ji, M.M.Musakhanov, A.M.Rakhimov
for usefull discussions.\\

}

\end{document}